\documentclass[apjl]{emulateapj}

\newcommand{\msun}{$M_{\odot}$}

\newcommand{\feh}{[Fe/H]}
\newcommand{\fcemp}{$f_{\rm CEMP}$}

\begin{document}
\title{Carbon-Enhanced Metal-Poor Stars, the Cosmic Microwave Background, and the Stellar IMF in the Early Universe}
\author{Jason Tumlinson}
\affil{Yale Center for Astronomy and Astrophysics, Yale University, P. O. Box 208121, New Haven, CT 06520}
\begin{abstract}
The characteristic mass of stars at early times may have been higher than today owing to the cosmic microwave background (CMB). This study proposes that (1) the testable predictions of this ``CMB-IMF'' hypothesis are an increase in the fraction of carbon-enhanced metal-poor (CEMP) stars with declining metallicity and an increase from younger to older populations at a single metallicity (e.g. disk to halo), and (2) these signatures are already seen in recent samples of CEMP stars and can be better tested with anticipated data. The expected spatial variation may explain discrepancies of CEMP frequency among published surveys. The ubiquity and time dependence of the CMB will substantially alter the reconstruction of star formation histories in the Local Group and early Universe.
\end{abstract}

\journalinfo{Accepted for publication in ApJ Letters}

\section{Introduction: The CMB and the IMF}

Just as the mass of a star determines its evolution and appearance, the stellar initial mass function (IMF) is a fundamental property of a stellar population. Because the IMF controls the energetic feedback and observational features of a stellar cluster, knowledge or ignorance of its form will shape the interpretation of galaxy data from high redshift and star formation histories reconstructed from surviving stellar fossils of early epochs.

The IMF, expressed by a typical or {\em characteristic} mass, $M_c$, may depend somehow on the metallicity of the star-forming gas. Both pure theory \citep{abn02,bcl02} and Galactic halo chemical abundances \cite[][hereafter T06]{tvs04,t06} indicate that the first generation (``Pop III'') was dominated by massive stars, $M_c \gtrsim 10 -  100$ \msun. To account for low-mass stars at [Fe/H] $\sim -4$, theorists have defined a ``critical metallicity'', $Z_{crit} \simeq 10^{-5.5} - 10^{-3.5} Z_{\odot}$ \citep{sch02,bl03,ss06,fjb07}, above which cooling and fragmentation are possible but where the metallicity dependence of the IMF may be complex \citep{omukai05}.

It is the influence of metals on the thermal evolution of star-forming gas, not the metals {\it per se}, that alters the IMF. At low metallicity cooling rates per mass decline and a cloud is typically warmer at a given density, which increases the fragmentation mass. Larson (1998; 2005) has explored the CMB as a heat source during early star formation and concluded that its influence could have a strong effect on the IMF in the first galaxies where $T_{CMB}$ will significantly exceed 10 K. Though high-$z$ signatures should be sought when access to $z > 6$ is routine, evidence for this ``CMB-IMF'' hypothesis can be pursued now using independent evidence preserved in ancient stars in the Local Group and recoverable using the tools of ``Galactic Archaeology''~\citep{fbh02}.

The goals of this study are: (1) to propose a link between the CMB and a new chemical signature of the IMF in the range $M = 1 - 8$ \msun\ - the carbon enhanced metal-poor (CEMP) stars, and (2) to propose observational tests that could support or falsify this hypothesis using data from upcoming surveys. CEMP stars are the subset of metal-poor stars ([Fe/H] $\leq -2.0$) that show elevated carbon relative to iron, [C/Fe] $> 1.0$ (Beers \& Christlieb 2005). The ratio of these stars to C-normal metal-poor stars has been linked to the IMF by models that explain the origin of the C enhancement in terms of mass transfer across a binary system \citep[hereafter K07,T07, respectively]{suda04, lucatello05b, komiya07, t07}. This origin scenario is used here to interpret the fraction of CEMP stars in a population relative to the total number, \fcemp, in terms of the IMF. These earlier papers may be consulted for more detailed discussion of the binary mass transfer model.

The CMB influences the IMF by establishing a temperature minimum, and therefore a characteristic fragmentation scale, for star-forming gas. \cite{rbl05} argued that the characteristic mass in local star-forming clouds ``is determined by the transition from an initial cooling phase of collapse to a later phase of slowly rising temperature that occurs when the gas becomes thermally coupled to the dust.'' A relation for $M_c$ as a function of the temperature where this reversal occurs, $T_{\rm min}$, begins with the Jeans mass:
\begin{equation}
M_J = \left( {\pi k T_{\rm min}} / {2 m_H G} \right)^{1.5} \rho^{-0.5}
\end{equation}
In the early cooling phase of collapse, the temperature-density relation can be expressed with a simple power law-dependence, $\rho = 10^{-18}(4.4K/T)^{3.7}$ g cm$^{-3}$, where $\rho _{\rm crit} = 10^{-18}$ g cm$^{-3}$ is the approximate density of turnaround in gas with $Z=Z_{\odot}$ and $T = 8$ K. Thus:
\begin{equation}
M_J = 4.4 \left( {T}/{10 \,{\rm K}} \right)^{\alpha} M_{\odot}
\end{equation}
where the exponent $\alpha = 3.35$. The Bonner-Ebert mass, which better reflects spherically symmetric collapse of an overdense fluctuation, is $M_{BE} = M_J / 4.7$. In the local ISM, star-forming clouds are observed to have $T_{min} \gtrsim 8$ K, so $M_{BE} = 0.44$ \msun, close to the observed peak of the local IMF (Kroupa 2002). Numerical simulations that have implemented power-law equations of state to study the effect of varying $T_{\rm min}$ and $\rho _{crit}$ on $M_c$ \citep{jklm05, ksj07} have found a shallower exponent, $\alpha \simeq 1.7$. Then, in terms of redshift:
\begin{equation}
\frac{M_{c}}{M_{\odot}} = \frac{M_{BE}}{M_{\odot}} = M_{norm} \left( \frac{max[2.73(1+z),8]}{10 \, {\rm K}} \right)^{\alpha}, \label{cmb-imf}
\end{equation}
where the coefficients $M_{norm} = 0.73$ and $M_{norm} = 1.06$ are obtained for $\alpha = 3.35$ or $1.7$, respectively, to recover the local value $M_c = 0.5$ \msun\ at 8 K. For definiteness, the IMF is specified with a universal log-normal form:
\begin{equation}
\ln \left( \frac{dN}{d\ln M} \right)= A - \frac{1}{2\sigma ^2}
       \left[ \ln \left( \frac{M}{M_c} \right) \right] ^2
\end{equation}
where $\sigma$ is the width and $A$ is normalization constant. The log-normal function offers flexible behavior with only one more free parameter than a power law.

These relationships are too simple to express the full complexity of the interplay of background radiation and metallicity in specifying $M_c$, but they are adequate to implement the CMB-IMF hypothesis easily within chemical evolution models. More detailed calculations are needed to work out the joint effects of varying metallicity, dust content, and background temperature in full detail, though such calculations already indicate that low-metallicity gas can cool to the CMB temperature at high redshift \citep[e.g.,][]{omukai05, smith07}\footnote{If the metal-poor IMF has the bimodal shape implied by multiple temperature minima in the cooling curves of Omukai et al. (2005), then this analysis applies only to the low-mass peak with $M_c \sim 1 - 10$ \msun.}. The next section shows that this simple CMB-IMF hypothesis explains a key observation of CEMP stars and predicts new testable consequences.

\section{The CMB-IMF Hypothesis and CEMP Stars}

CEMP stars are thought to be metal-poor low-mass stars ($M_2 \lesssim 0.8$ \msun) that have acquired C enhancements at their convective surfaces by capturing the C-rich ejecta of an AGB companion ($M_1 = 1.5 - 8$ \msun). They represent approximately 20\% of Population II stars at [Fe/H] $< -2$, $\sim 40$\% at [Fe/H] $< -3.5$, and all three stars known with [Fe/H] $< -4.5$ are CEMP \citep{he0107,he1327,he0557}. The case for a binary origin of most CEMP stars is based on two observations: (1) $\simeq 80$\% show high enhancement of barium ([Ba/Fe] $>$ 1), a ``main s-process'' element that strongly indicates an AGB origin \citep{busso99, aoki07}, and (2) these ``CEMP-s'' stars show radial velocity variations often enough to be consistent with 100\% binarity \citep{lucatello05a}. However, the 20\% of CEMP stars that are not s-enhanced (``CEMP-no'') are ambiguous: they may arise in binary systems with an AGB primary of $M_1 = 5 - 8$ \msun\ that did not produce or dredge up s-process elements before the mass transfer (K07) or they may have acquired their light-element enhancements from another mechanism that is unrelated to binarity, such as an unusual supernova. To mitigate this uncertainty, this study corrects the observed frequency of all CEMP stars by a factor 0.8 to exclude the CEMP-no stars.

Because the binary circumstances that produce a CEMP require both a low-mass star and an intermediate mass star, the incidence of CEMP stars reflects the underlying IMF in the range $ 1 - 8$ \msun\ (K07, T07). Here I assume that the IMF describes the cloud core mass $M = M_1 + M_2$, that a fraction $f_b = 0.6$ of all stars $ < 8$ \msun\ form in binaries, that all binary mass ratios $q = M_1/M_2$ are equally likely but $q > 0.1$ \citep{dm91}, and that half of IMS/LMS binaries give a CEMP outcome \citep{lucatello05b}. These assumptions are reasonable compared to local stellar populations (K07,T07).


The CMB introduces a time dependence to the IMF (Eq.~\ref{cmb-imf}). When coupled to chemical evolution models, this behavior yields two predictions that can test the CMB-IMF hypothesis. First, chemical evolution proceeds locally in the galaxy hierarchy such that as a general rule a given region will increase in metallicity over time. Coupling this trend to the CMB-IMF hypothesis gives the testable expectation that the fraction of CEMP stars, \fcemp, in a population should increase with declining metallicity (\S~3.1). Second, the hierarchical nature of galaxy formation segregates different star forming regions from one another during galaxy formation, and their chemical evolution proceeds at different rates. Though spatial variations are partially damped by mixing and feedback, chemical evolution is local and stars at the same metallicity can form at different times. This trend, when added to the CMB-IMF hypothesis, makes the prediction that \fcemp\ should vary spatially at the same metallicity (\S~3.2), increasing in older populations and decreasing in younger ones at fixed [Fe/H]. The next sections refine and tests these two predictions.

\subsection{Variation with Metallicity}
\label{results1}

\begin{figure}
\plotone{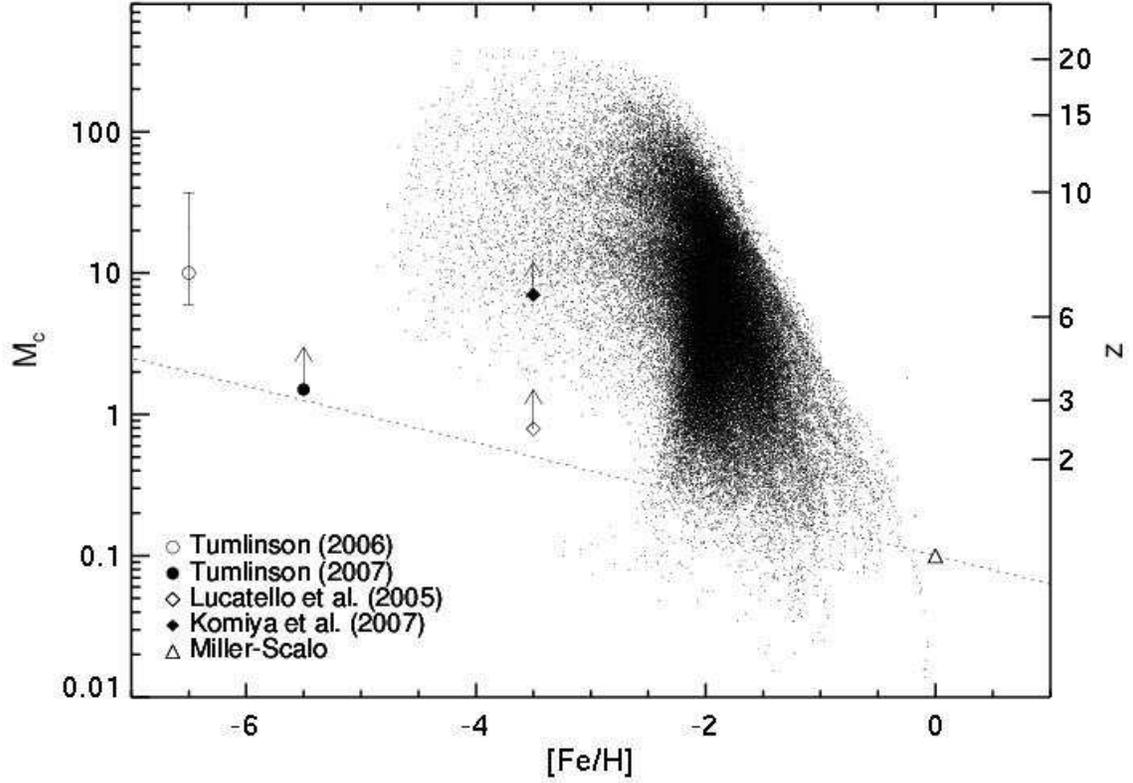}
\caption{Chemical abundance constraints on $M_c$ compared to models incorporating the CMB-IMF hypothesis. The Pop III $M_c$ from T06 are plotted arbitrarily at [Fe/H] $= -6.5$. Small points mark chemical trajectories for a fiducial model of the Milky Way halo from T06 with $\alpha = 3.35$. There is a general trend to higher metallicity over time but also scatter in formation time at a single [Fe/H]. The redshifts at right are converted from $M_c$ using Eq.~\ref{cmb-imf} and $\alpha = 3.35$. The $\sim 10^6$ points in the qualitatively similar pattern for $\alpha = 1.7$ are omitted for clarity. \label{fig1}} \end{figure}

Chemical evolution is a generally local phenomenon in which the metal content of a region increases with time. Temporary reverses can follow accretion of metal-poor gas or blowout by supernovae, but these are second order effects on the broad trend. Figure~\ref{fig1} shows the time evolution of metallicity and $M_c$ for a fiducial stochastic halo model from T06 with $alpha = 3.35$. The general trend toward higher metallicity at later times is evident even when many stochastically evolving subhalos are combined. Here, the earliest subhalo begins forming stars at $z \sim 20$ and quickly becomes metal-enriched. This model includes a primordial IMF with $M_c = 10$, $\sigma =1.0$ (T06 case A). The composition of primordial stars, not the CMB, sets their IMF at $z \lesssim 70$ ($\sim 200$ K, where H$_2$ cooling plateaus). Though their yields may influence the mainstream $\alpha$ and Fe-peak abundances in the CEMP stars, primordial stars are unlikely to cause \fcemp\ $\sim 0.4$ at [Fe/H] $\sim -3$. Stars with [Fe/H] $\sim -3$ form until $z \simeq 4$ but generally earlier than those of [Fe/H] $\sim -2$ which continue until $z \simeq 1$. This rise with time coupled to the CMB-IMF implies higher $M_c$ at high $z$ and an increase in \fcemp\ with declining [Fe/H]. Figure~\ref{fig1} compares these histories with observational constraints from CEMP survey \citep[][K07,both lower limits]{lucatello05b} and the hyper-metal-poor (HMP) stars (T07), which are met by the CMB-IMF hypothesis.

\begin{figure}
\plotone{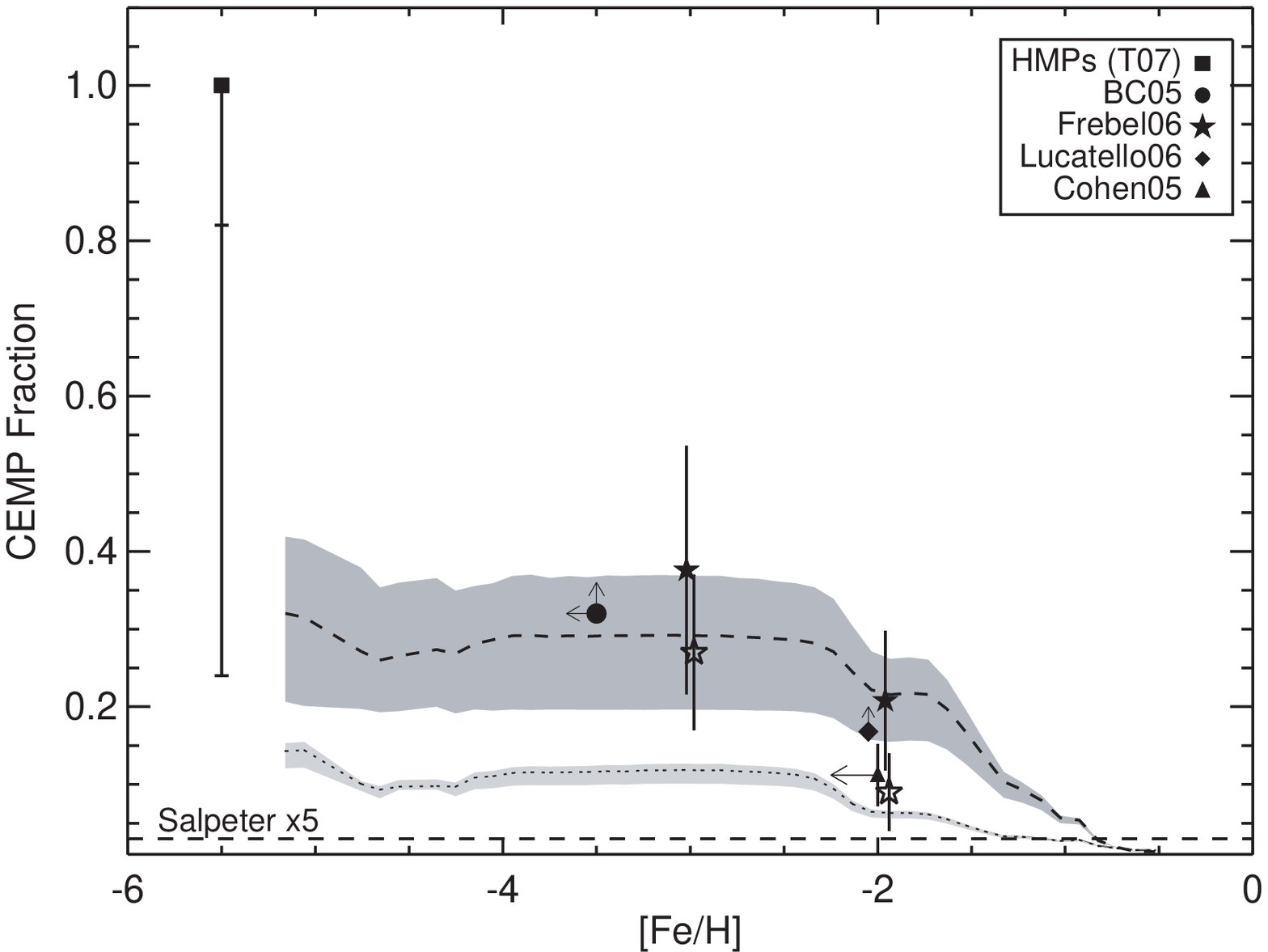}
\caption{Variation of \fcemp\ with \feh\ in data and models. Value of \fcemp\ are calculated with $\alpha = 1.7$ (dotted) and $3.35$ (dashed) and the $M_1 = 1.2 - 5.0$ \msun\ fiducial primary mass range, with the shaded regions taking in $M_1 < 3.5$ to $8.0$ \msun. Observational estimates are shown from the HMPs (T07, with 1 and 2 $\sigma$ limits) to [Fe/H] $\simeq -2$ (corrected by 0.8). For the \cite{frebel06} sample, the cumulative fraction above two thick disk scale heights (3 kpc) is used. Also shown is $5f_{CEMP}$ for a Salpeter IMF. \label{fcemp-vs-feh-fig} } \end{figure}

Figure~\ref{fcemp-vs-feh-fig} compares \fcemp\ from data and models at varying [Fe/H]. The concentration of [Fe/H] $\lesssim -3$ stars at $z > 4$ places them at \fcemp\ $\simeq 0.2 - 0.4$, in the range of the observations. Stars with [Fe/H] $\simeq -2$, near the Galactic halo mean, form down to $z \simeq 1$ and so lie in a transition region from low to high \fcemp, while the typical surviving [Fe/H] $\simeq -1$ star forms at $z < 2$ in a nearly modern IMF. Thus the CMB-IMF hypothesis embedded in a realistic model of Galactic chemical evolution can explain the increase in \fcemp\ at low [Fe/H].

The main uncertainties in this model are the power-law dependence of $M_c$ on temperature, for which two extremes of $\alpha$ are calculated, and the uncertainty in the most massive AGB primary star that can give C enhancement. Lower values of $\alpha$ apparently fail to match \fcemp\ unless more than 50\% of binaries give a CEMP or better statistics in the outer halo bring down the \fcemp\ measurements. The fiducial model follows K07 in setting $M_1 = 1.2 - 5.0$ \msun. AGB stars of $5-8$ \msun\ may be eligible to donate C to a secondary, but these stars should show N enhancement from hot-bottom burning (HBB) during dredge-up in more massive AGB stars \citep{herwig05}. That N-rich stars are uncommon \citep{jj07} may indicate that this mass range either does not experience HBB or does not form in close binaries. To cover this uncertainty, the shaded regions in Figure~\ref{fcemp-vs-feh-fig} display models for AGB primary masses ranging from 1.2 \msun\ to 3.5, 5.0, and 8.0 \msun. Binaries with $M_1$ in the excluded range are included in the denominator of \fcemp, but not counted as CEMP stars. This uncertainty is roughly a factor of two in the $\alpha = 3.35$ model.


\subsection{Variation with Location in the Galaxy}
\label{results2}

To couple chemical evolution to the underlying dark matter dynamics, the T06 merger-tree based stochastic chemical evolution model is being adapted to work within merger trees extracted from N-body models of Milky Way assembly. This new method allows for calculations of spatial variations of CEMP fractions when added to the CMB-IMF hypothesis.

\begin{figure}
\plotone{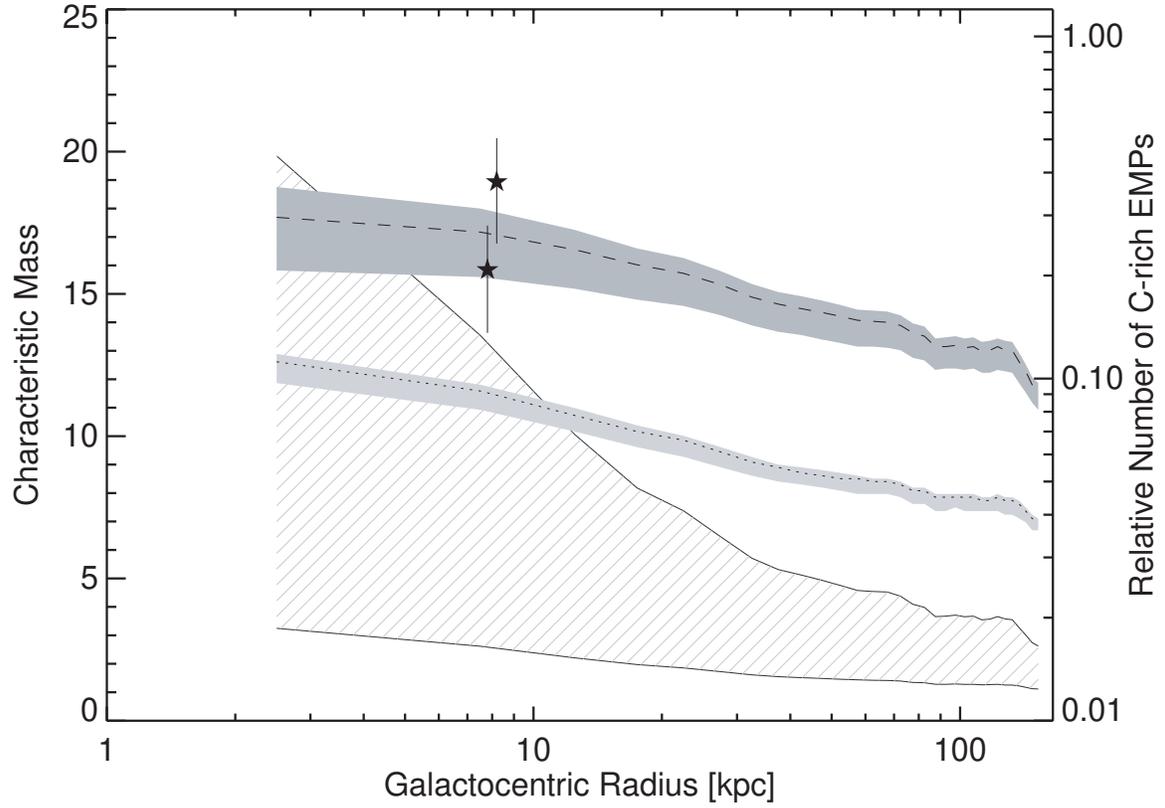}
\caption{Run of characteristic mass $M_c$ (cross-hatched) and expected \fcemp\ (shaded, same meanings as Figure 2) in the most metal-poor populations, versus Galactocentric radius. The expectation of \fcemp\ $\simeq 0.2 - 0.4$ in the halo at the solar circle agree well with the findings of \citet[][filled stars]{frebel06} for bright HES stars more than two thick-disk scale heights (3 kpc) above the midplane and [Fe/H] $ = -3$ to $-2$. \label{mc-fcemp-vs-radius-fig} } \end{figure}

First, realizations of the Galactic dark-matter assembly history are created using the Gadget TreePM code \citep[][version 2.0]{gadget2} with $256^3$ particles in a $125$ comoving Mpc$^3$ cubic box and a WMAP3 cosmology \citep{wmap3}. Snapshots generated at $\Delta z \sim 0.1$ intervals are searched for virialized halos of $\geq 32$ particles ($M_{DM} \geq 1 \times 10^{7}$ \msun) using a friends-of-friends algorithm. For each particle in the present-day halo of $M_{DM} \sim 2 \times 10^{12}$ \msun, I determine when that particle first entered a virialized object that forms a part of the final halo. This redshift is converted into $M_c$ (Eq.~\ref{cmb-imf}), assuming that stars enrich up to [Fe/H] $ \simeq -3$ promptly after virialization (see Figure~\ref{fig1}). The central condensation of the earliest substructure is a generic feature of CDM models of galaxy formation, so that the oldest stars preferentially reside near the center of the final halo \citep{diemand05,es06}. This gradient of formation time translates into a gradient in $M_c$ and \fcemp\ with radius (Figure~\ref{mc-fcemp-vs-radius-fig}), which can be tested by appropriately selected CEMP surveys.

There is already tentative evidence that \fcemp\ varies spatially in the Galaxy at the same metallicity: \citet[ Figure 11]{frebel06} find \fcemp\ $= 0.09$ for [Fe/H] $\leq -2.0$ in the Galactic midplane, increasing to $0.27 - 0.67$ at Galactic $Z = 3-6$ kpc, and \fcemp\ $= 0.27$ in the midplane for [Fe/H] $\leq -3.0$, increasing to $\sim 0.46 - 1.0$ at $Z = 3-6$ kpc (the higher Z bins have few stars and are averaged for Figures 2 and 3). This behavior is difficult to understand if $M_c$ is influenced solely by metallicity. \cite{frebel06} did not distinguish halo from thick disk stars; the latter probably dominate at $Z < 3$ kpc because the sample was selected for brightness (T. C. Beers 2007, private communication). Thus a full kinematic analysis of this sample and others with better statistics in the outer halo will be required to test the CMB-IMF hypothesis. Nevertheless, this trend provides the first tentative evidence that \fcemp\ varies with location at a single metallicity. This behavior is predicted by the CMB-IMF hypothesis but should not occur if the IMF has only a metallicity dependence.  The predicted spatial variation may also explain discrepancies in \fcemp\ from study to study, which currently range from 0.09 \citep{frebel06}, to 0.14 \citep{cohen05}, to $> 0.21$ \citep{lucatello06}. This discrepancy might result if these samples preferentially cover different regions of the Galaxy with intrinsic spatial variation introduced by the CMB. It is interesting that this disagreement may result from a physical cause and not systematic errors. These trends could all be tested by kinematically selected, unbiased samples of CEMP stars that carefully distinguish older from younger stellar populations.

\section{Discussion and Conclusions}

In summary, the CMB-IMF explains several puzzling observational facts within a single physical model with strong independent justification. The variation of CEMP frequency with metallicity and possible spatial variation in the halo provide tentative evidence that the CMB controls the characteristic stellar mass during the early phases of galaxy formation. These effects appear in a Milky Way chemical evolution model that includes the CMB-IMF in a simple parametric form. Though the connection between the CMB and CEMP stars is still speculative and includes some likely oversimplifications, it has observable consequences that can be tested with upcoming large surveys (e.g., SDSS-SEGUE). This model also suggests an explanation for discrepancies in published measurements of CEMP fraction.

The CMB-IMF model has three uncertain features that require additional study. The first is that the theoretical form of the IMF with background temperature, expressed by $\alpha$, is unknown. This issue should be addressed by numerical simulations to examine the response of the fragmentation mass scale to joint variation of metallicity and CMB temperature. Second, the model adopts properties of binary stars from Galactic disk conditions that need not hold at low metallicity, though early data suggest they do. Fortunately these properties are measurable directly from the data that will test the CMB-IMF with \fcemp. Third, the model relies on a specific but uncertain relationship between stellar mass and AGB C and s-process nucleosynthesis. Though the model will tolerate variations in the mass range AGB stars that give the s process (see \S~3.1), a radical revision of our understanding of the s-process site could invalidate the binary origin for CEMP-s stars and break their link to the IMF and the CMB. However, it is more likely that continuing theoretical refinements in AGB models will improve the mapping from CEMP stars to the IMF. These issues will be explored in subsequent publications, along with further tests and implications of the CMB-IMF.

The CMB-CEMP connection demonstrates the potential of ``Galactic archaeology'' to complement high-$z$ studies with new discoveries about star formation in the early Universe. \cite{rbl98} and \cite{chabrier03} have discussed some implications of a non-standard IMF, but the unique status of the CMB implies at least two others that merit further examination. First, a CMB-IMF will cause systematic {\em underestimates} of early star formation rates in color-magnitude reconstructions that assume a normal IMF \citep{brown06, cole07} and systematic {\em overestimates} of star formation rates from the rest-frame UV light of massive stars at high $z$ \cite[e.g.,][]{labbe06}. Second, a CMB-IMF could explain the extreme mass-to-light ratios ($\sim 100 - 500$) observed in the Milky Way's faintest satellites \citep{willman06,belo07}. If these formed at high redshift the relatively few remaining low-mass stars should show high \fcemp. Thus a CMB-IMF may significantly affect our emerging understanding of how galaxies form and evolve beyond the current $z \sim 7$ frontier.

\acknowledgements
This paper is dedicated to the late Gilbert Mead and to Jaylee Mead in gratitude for their support of the Mead fellowship in YCAA. Talks with R. Larson, P. Coppi, T. Beers, J. Johnson, M. Pinsonneault, M.-M. Mac Low, and A. Szymkowiak have improved the paper. I am also grateful to Achim Weiss for his thorough and constructive refereeing.


\clearpage

\clearpage

\clearpage

\end{document}